# Jet propagation velocity and environment density of giant radio sources with steep radio spectrum

Alla P. Miroshnichenko

Institute of Radio Astronomy, National Academy of Sciences, Ukraine

e-mail: mir@rian.kharkov.ua

**ABSTRACT**

Before we have determined that giant radio structure is typical for steep-spectrum sources detected with Ukrainian T-shaped Radio Telescope of model 2 (UTR-2). To estimate the jet propagation velocity and environment density of giant radio sources we consider the sample of galaxies and quasars with steep radio spectrum compiled from two regions of UTR-2 catalogue.

The value of velocity of jet propagation is obtained from our estimates of linear size of radio structure and characteristic age of a source. We use the obtained values of jet propagation for the estimates of density of jet environment, that is, density of environment at galaxies and quasars outskirts.

We have derived that values of jet propagation velocities of examined sources are $\sim 10^9$ cm/s. The positive correlation of jet velocity and redshift of source is displayed in our sample. On the assumption of equality of jet luminosity and corresponding kinetic luminosity we estimate the density of jet environment (radio structure environment) for galaxies ($10^{-26}$ g/cm$^3$ – $10^{-27}$ g/cm$^3$) and quasars ($10^{-28}$ g/cm$^3$). Considered steep-spectrum galaxies and quasars display strong evolution of jet propagation velocity and jet environment density.

Properties of giant steep-spectrum sources are in accordance with conception of long evolution of powerful sources.

The steep low-frequency spectra may be as markers of giant radio structures of galaxies and quasars.

## 1 INTRODUCTION

The catalogue UTR-2 is the catalogue of extragalactic radio sources detected at frequencies 10, 12.6, 14.7, 16.7, 20 and 25 MHz during the northern sky survey at the largest over the world Ukrainian radio telescope UTR-2 in the decametre band (Braude et al., 1978, 1979, 1981, Konovalenko et al., 2016). Radio spectra of detected sources in UTR-2 catalogue had been derived at the range 10 – 1400 MHz with connection of more high-frequency data and had shown that near 30 per cent of low-frequency spectra are steep (the value of spectral index was greater than 1) (Braude et al., 2003). At that, near 10 per cent of these steep spectra have a break at low frequencies with spectrum rising at the decameter band. Earlier, this spectral type named as C+ was noted in the results of observations with the Ukrainian decameter radio telescope of the first generation UTR-1 (Braude et al., 1970). In the same time the quantity of linear type of steep radio spectra (type S) is near twice the quantity of steep low-frequency spectra (Miroshnichenko, 2012). One can see at the Fig.1 the samples of radio spectra with types linear S (steep and non-steep) and break C+ for some sources from UTR-2 catalogue.

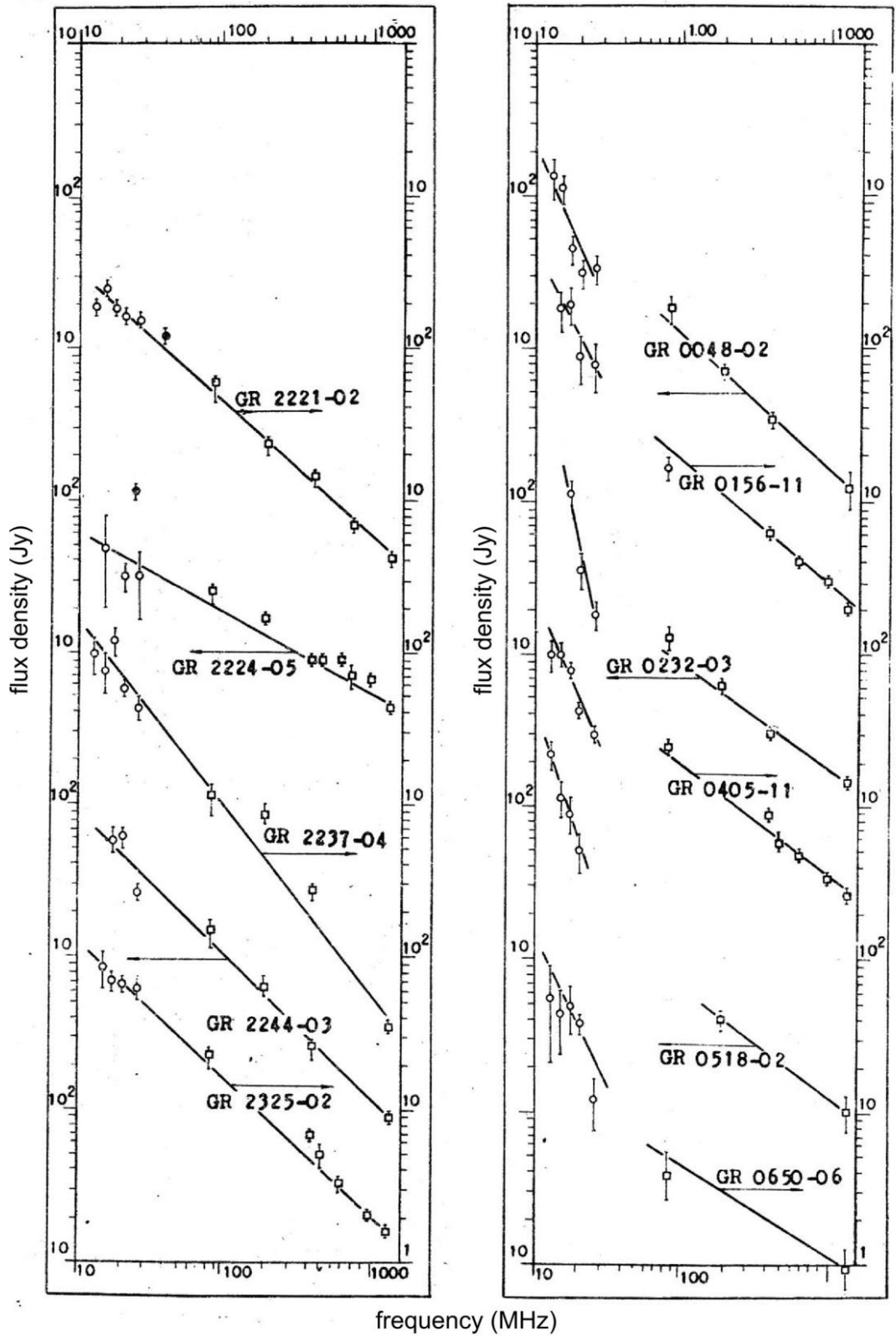

Figure 1. Examples of linear (S-type) and break (C+ type) of some radio spectra of sources from the UTR-2 catalogue (from (Braude et al. 1981))

Before, we noted (Miroshnichenko, 2010) that observed low-frequency radio spectra are in according with known Kardashev mechanism of energy losses of relativistic electrons at the transient injection of relativistic particles in the source (Kardashev, 1962). In this case the value of the spectral index changes from the standard value $\alpha = (\gamma - 1)/2$ to $\alpha = (2\gamma + 1)/3$, where $\gamma$ is the index of the power distribution of the energy of relativistic electrons (Kardashev, 1962). Since the critical frequency of synchrotron radiation of relativistic particles decreases with time, its value may located at frequencies less than 10 MHz. It is known, the source decametre emission at the synchrotron mechanism corresponds to "oldest" relativistic electrons having essentially low energies in comparison with those radiating at high-frequency band. So, the observations of decametre emission of radio sources have advantage at the determination of many astrophysical characteristics of objects and their evolution (Konovalenko et al. 2016).

The particular interest is excited by the radio sources with steep low-frequency spectrum (the spectral index value is greater than 1). This spectrum steepness may be reflection of the specific long evolution of given sources. As we have obtained (Miroshnichenko, 2010, 2012, 2013, 2015) the properties of steep-spectrum sources from the UTR-2 catalogue (Braude et al., 1978, 1979, 1981a, 1981b, 2003) are peculiar: all galaxies and quasars with steep radio spectrum have high luminosity (the monochromatic radio luminosity at 25 MHz is $\sim 10^{28}$ W/(Hz ster)), giant linear size of radio image ($\sim$ Mpc), great characteristic age ($\sim 10^8$ years). Giant size of radio structure of these sources is connected with great extent of jets emanated from active nuclei of objects. We consider that such extent may indicate on the great velocity of jet propagation. Besides, the low density of extragalactic medium may provide the jet propagation on the Mpc – scales. Thus, the aim of this work is determination of important physical parameters of sources with low-frequency steep spectrum – their jet propagation velocity and environment density.

## 2 JET PROPAGATION VELOCITIES OF STEEP-SPECTRUM SOURCES

In order to estimate the jet propagation velocity and the density of their environment we consider the sample of galaxies and quasars with low-frequency steep radio spectrum compiled from two regions of UTR-2 catalogue (declinations are from –13 degrees to +20 degrees and from +30 degrees to 40 degrees (Miroshnichenko, 2012, 2013, 2014)). At the selection criteria (the spectral index value at the decametre band is greater than 1 and the flux density of radio emission at 25 MHz is $S_{25} > 10$ Jy) we have identified 130 galaxies and 91 quasars with giant radio structures. The NED database has been used for optical identifications and angular sizes of examined objects. All calculations of physical characteristics of objects (linear size, monochromatic luminosity, magnetic field strength, characteristic age) have been held at the cosmological parameters $\Omega_m = 0.27$, $\Omega_\lambda = 0.73$, $H_0 = 71$ km/(s Mpc) (Miroshnichenko, 2010, 2012, 2013, 2014). Our sample displays the sources with two types of steep low-frequency radio spectra: type S (linear steep spectrum) and type C+ (break steep spectrum), including 78 galaxies with type S ($G_S$), 52 galaxies with type C+ ($G_{C+}$), 55 quasars with type S ($Q_S$), and 36 quasars with type C+ ($Q_{C+}$). The mean values of spectral indices are similar for examined galaxies and quasars of type S : $<a> = 1.21$ (+-0.02) for $G_S$ and $<a> = 1.24$ (+-0.03) for $Q_S$. Also, corresponding values are close for examined galaxies and quasars of type C+ : $<a> = 2.01$ (+-0.08) for $G_{C+}$ and $<a> = 2.25$ (+-0.14) for $Q_{C+}$. At that the mean values of spectral indices for corresponding high-frequency range of C+ spectra for examined galaxies and quasars are 0.74 (+-0.03) and 0.62 (+-0.05), respectively. It is noteworthy, that mean values of flux densities of steep-spectrum sources at the decametre band are not weak. For example, the mean flux density at frequency 25 MHz is near 43 Jy for galaxies and near 36 Jy for quasars.

Angular sizes of sample's sources were estimated from NVSS radio images derived at frequency 1.4 GHz with interferometer VLA (at angular resolution of 45" and sensitivity = 0.45 mJy/beam) (Condon et al., 1998). We suppose that the frequency changing of angular sizes of sources at the frequency range from 1.4 GHz to 25 MHz is not essential. From examination of the NVSS maps we have found that sources with steep low-frequency spectrum are practically not affected by the confusion effect (Miroshnichenko, 2012). The analysis of the NVSS maps included vicinities of the sources with steep linear spectra S-type and with break spectra C+. For one, the examined NVSS map with considered source from UTR-2 catalogue GR 0853+12 (4C 12.32) contains (at the field of radius 0.5 degrees ) more 3 sources with comparative intensity to 4C12.32. At that the source 4C 12.32 has steep linear spectrum S with spectral index $a = 1.28$. Besides this source had been examined by Gopal-Krishna et al. (2005). They pointed out the same value of the spectral index for 4C 12.32 ($a = 1.28$) (Gopal-Krishna et al., 2005). If the steepness of low-frequency spectra is due to confusion, then sources with steep linear spectrum S would not observed, and only sources with break spectrum C+ must be observed. But UTR-2 catalogue displays almost twice as many sources with steep S-spectra, than sources with C+ spectra.

It is of importance, that orientation of the jet axis with respect to the observer's line-of-sight as 90 degrees is assumed for all giant sources (Machalski et al., 2004). Hence, we can calculate linear sizes of these sources without account of projection effect.

The value of velocity of jet propagation $V_j$ is obtained from the estimates of linear size of radio structure $R$ and characteristic age of source $t$ (Miroshnichenko, 2005) :

$$V_j = R / 2\, t \qquad (1)$$

In this paper we use our previous estimates of *R* and *t* for examined galaxies and quasars (Miroshnichenko, 2012, 2013). Note, the characteristic age of source *t* was determined by the spectral ageing method (Alexander & Leahy, 1987; Daly, 1995; Jamrozy et al., 2005). In this case the characteristic age *t* is equal to characteristic time of synchrotron radiation losses $t_{syn}$ in the source's magnetic field (e. g. Jamrozy et al., 2005) :

$$t_{syn} = 50{,}3 \frac{B^{1/2}}{B^2 + B_{CMB}^2} [\nu_b(1+z)]^{-1/2} \qquad (2)$$

where $t_{syn}$ is the characteristic age, $10^6$ yr, *B* is the source's magnetic field strength, $10^{-5}$ G, $B_{CMB} = 0{,}32(1+z)^2$ is the strength of the equivalent magnetic field corresponding to the intensity of the microwave background, $10^{-5}$ G, $\nu_b$ is the critical frequency of the synchrotron spectrum, GHz. The estimates of $t_{syn}$ have been made for a critical frequency $\nu_b$ = 10 MHz (Miroshnichenko, 2012).

Also, some authors employ the method of dynamical age $t_{dyn}$ for estimates of source's ages (Kaiser & Alexander, 1997; Kino & Kawakatu, 2005; Machalski et al., 2008). This method needs modeling of jet power, energy density and pressure in the radio lobes, the total energy of the source. It is commonly accepted assumption about the proportionality of the spectral and dynamical ages (up to $t_{dyn}$ = 2 $t_{syn}$ ) (Machalski et al., 2004).

It turns out from (1) that the values of jet propagation velocities of sample's sources are ~ $10^9$ cm/s. So, these are subrelativistic both for galaxies and quasars with steep radio spectra at given four subsamples: $G_S$ , $G_{C+}$ , $Q_S$ , $Q_{C+}$. This conclusion follows from the obtained mean values of jet propagation velocity < $V_j$ > for examined classes of sample's objects:

$G_S$            < $V_j$ > = $2.97*10^9$ (+- $0.67*10^9$) cm/s ;

$G_{C+}$          < $V_j$ > = $5.78*10^8$ (+- $4.09*10^8$) cm/s ;

$Q_S$            < $V_j$ > = $3.12*10^9$ (+- $0.31*10^9$) cm/s ;

$Q_{C+}$          < $V_j$ > = $1.65*10^9$ (+- $0.45*10^9$) cm/s

Note, that 10 per cent of examined sample's galaxies with steep radio spectrum are members of galaxy clusters. We have obtained for these galaxies the mean velocity value

     < $V_j$ >$_{cluster}$ = $5.64*10^8$ (+- $1.43*10^8$) cm/s,

separately including for galaxies with spectral type S one is

     < $V_j$ >$_{cluster}$ = $7.19*10^8$ (+- $1.55*10^8$) cm/s,

and for galaxies with spectral type C+ one is

     < $V_j$ >$_{cluster}$ = $4.94*10^7$ (+- $0.51*10^7$) cm/s.

Thus, the values of jet propagation velocity of sample's cluster galaxies are smaller than these for other sample's galaxies. Taking into account that cluster environment is denser than extragalactic environment we can see the slowing of the jet propagation in galaxy clusters.

Galaxies and quasars of our sample cover the wide range of redshifts z (from z = 0.006 to z = 3.57), allowing the study their derived parameters versus the redshifts. All four classes of examined sources display the positive correlation of jet propagation velocity and redshift (see Fig. 2-5):

$G_S$   $V_j \sim (1+z)^{2.71\ (+-\ 0.09)}$ ,

$G_{C+}$   $V_j \sim (1+z)^{5.23\ (+-\ 0.21)}$ ,

$Q_S$   $V_j \sim (1+z)^{2.31\ (+-\ 0.13)}$ ,

$Q_{C+}$   $V_j \sim (1+z)^{4.69\ (+-\ 0.29)}$ .

As one can see from determined relations (Fig. 2-5), the jet propagation velocity for sources with steep radio spectrum is greater for more early cosmological epochs and it decreases to small redshifts by power law. At that the more prominent evolution of the jet velocity is observed for examined objects with spectral type C+ ($G_{C+}$ and $Q_{C+}$). This may be due to an activity recurrence of nuclei of sources with steep spectrum C+.

## 3 ESTIMATES OF JET ENVIRONMENT DENSITIES OF STEEP-SPECTRUM SOURCES

We use the obtained values of the jet propagation velocities $V_j$ for the estimates of the jet environment densities, that is, density of environment at galaxies's and quasars's outskirts in given sample. Note the linear sizes of examined objects have values of about Mpc (Miroshnichenko, 2012). To calculate the densities we assume the equality of jet luminosity $L_j$ and corresponding kinetic luminosity $L_k$ (Leahy, 1991):

$$L_j = \frac{4}{3}\pi r_j^2 V_j U \qquad (3)$$

$$L_k = \frac{\pi}{2} r_j^2 \rho_j V_j^3 \qquad (4)$$

where $r_j$ is the jet radius, $U = \frac{7}{3}\frac{B^2}{8\pi}$ is the minimal energy density of a source, $B$ is the magnetic field strength of a source, $\rho_j$ is the jet density.

It follows from (3) and (4) that the jet density (environment density) $\rho_j$ has value :

$$\rho_j = \frac{7B^2}{9\pi Vj^2} \qquad (5)$$

From (5) we get the estimates of density of jet environment $\rho_j$ for galaxies and quasars with steep radio spectra. For examined galaxies these values are from $10^{-26}$ g/cm$^3$ to $10^{-27}$ g/cm$^3$, and for examined quasars these are of about $\sim 10^{-28}$ g/cm$^3$. The mean values of jet environment density $<\rho_j>$ for each class of sample objects are next:

  Gs  $<\rho_j> = 1.32*10^{-27}$ (+- $0.36*10^{-27}$) g/cm$^3$ ,

  Gc+  $<\rho_j> = 1.30*10^{-26}$ (+- $0.32*10^{-26}$) g/cm$^3$ ,

  Qs  $<\rho_j> = 6.29*10^{-29}$ (+- $1.32*10^{-29}$) g/cm$^3$ ,

  Qc+  $<\rho_j> = 1.08*10^{-28}$ (+- $0.43*10^{-28}$) g/cm$^3$

Also, we obtain the environment density $\rho_{j\,cluster}$ for the sample's galaxies belonging to galaxy clusters (their number is near 10 per cent of all sample's galaxies). It turns out that this value is greater than one for isolated sample's galaxies. The mean value of environment density for cluster galaxies is

  $<\rho_j>_{cluster} = 6.80*10^{-27}$ (+- $2.83*10^{-27}$) g/cm$^3$.

In particular, this value is

  $<\rho_j>_{cluster} = 3.19*10^{-27}$ (+- $1.42*10^{-27}$) g/cm$^3$

for galaxies with spectral type S, and one is

  $<\rho_j>_{cluster} = 1.89*10^{-26}$ (+- $0.91*10^{-26}$) g/cm$^3$

for galaxies with spectral type C+.

  Thus, the jet environment of galaxies with steep low-frequency radio spectrum (especially for type C+) is denser than the jet environment of quasars. It is interesting to consider the relation of jet environment density and redshift of examined sources (see Fig. 6-9):

  Gs  $\rho_j \sim (1+z)^{-6.28\,(+-\,0.25)}$ ;

  Gc+  $\rho_j \sim (1+z)^{-11.49\,(+-\,0.87)}$ ;

  Qs  $\rho_j \sim (1+z)^{-5.10\,(+-\,0.32)}$ ;

  Qc+  $\rho_j \sim (1+z)^{-9.54\,(+-\,1.04)}$ .

As one can see, the sample's galaxies and quasars with steep radio spectrum of type C+ reveal the essential evolution of jet environment density (Fig. 6-9).

# 4 COMPARISON OF PROPERTIES OF SOURCES WITH STEEP AND NON-STEEP SPECTRA

We have examined properties of radio sources with ordinary spectra (that is, non-steep spectra) and steep spectra (Miroshnichenko, 2004, 2007) basing on samples compiled from data at high radio frequencies (Lawrence et al., 1986, Murphy et al., 1993, Reid et al., 1995, Lara et al., 2001).

**I**. For example, we derived from data at 1,4 GHz and 5 GHz by Lara et al. (2001) **for giant radio galaxies** their mean values of such physical parameters : linear size

$<R> = 2.05 (+-0.15) 10^{24}$ cm,

characteristic age

$<t> = 7.83 (+-0.68) 10^6$ yr,

jet propagation velocity

$<V_j> = 6.2 (+-0.31) 10^8$ cm/s,

spectral index

$<a> = 0,89 (+-0.10)$.

**II**. From data by Lawrence et al. (1986) we obtained **for moderate size sources** :

$<R> = 4.04 (+-0.43) 10^{23}$ cm,

$<t> = 6.53 (+-3.98) 10^5$ yr,

$<V_j> = 2.35 (+-0.85) 10^9$ cm/s,

$<a> = 0,70 (+-0.05)$.

**III.** From data by Murphy et al., 1993, Reid et al., 1995) we determined **for compact sources**:

$<R> = 3.3 (+-0.33) 10^{23}$ cm,

$<t> = 1.6 (+-0.24) 10^5$ yr,

$<V_j> = 1.68 (+-0.16) 10^9$ cm/s,

$<a> = 0,22 (+-0.06)$.

Thus, one can see the greater values of linear sizes, characteristic ages, spectral indices, but smaller values of jet propagation velocity for giant sources, than these for ordinary sources.

A number of authors (Carilli et al., 1991, Daly, 1995, Smith et al., 2002, Kino & Kawakatu, 2005) investigated some powerful radio sources, in particular, Cyg A (3C 405) which embedded in a galaxy cluster. They derived estimates of ambient gas densities (~ $10^{-27}$ g/cm$^3$), jet velocities (~ $10^{-2}$ c), characteristic age ( from 3 Myr to 30 Myr). Machalski et al. (2008) have estimated very low environment density (~ $10^{-31}$ g/cm$^3$), jet propagation velocity $V_j = (0.163 (+-0.021)$ c) and characteristic age ~ 47.6 Myr for giant galaxy J 1420-0545 with linear size of radio structure ~4.7 Mpc. From these results the authors conclude about large inhomogeneity of the intergalactic medium. LOFAR study of galaxy 3C 31 (Heesen et al., 2018) reveal the giant linear size of this source (1.1 Mpc) from its radio images at 52 MHz and 145 MHz. The authors obtained profiles of spectral indices in radio tails (lobes) where the spectra steepen with increasing distance from the nucleus. In particular, the spectral index between 52 and 145 MHz has value 1.2 – 1.4 at the distant ends of radio lobes. Note, the source 3C 31 belonging to our sample, displays that its linear size is ~ 1 Mpc , the value of spectral index between 10 MHz and 25 MHz is 1.8 (+-0.3).

## 5 CONCLUSIONS
+

At continuation of the examination of sources with low-frequency steep spectrum from the UTR-2 catalogue we have derived some important physical features of giant radio sources:

*Properties of giant low-frequency steep-spectrum sources are in accordance with conception of long evolution of powerful sources.

*Jet propagation velocities of giant steep-spectrum radio sources are subrelativistic (~ 0.1 light velocity). These velocities decrease from early cosmological epochs to present epoch due to losses of energy by relativistic particles in jets.

*Galaxies and quasars with break steep spectrum (type C+) have more strong evolution of jet propagation velocity than these with linear steep spectrum (type S). This may indicate on the nucleus activity recurrence for objects with steep radio spectrum C+.

*Galaxies and quasars with steep radio spectra of types C+ and S display strong evolution of their jet environment density. At that the value of jet environment density is greater for galaxies, especially, for galaxies of spectral type C+.

*The steep low-frequency spectra may be as markers of giant radio structures of galaxies and quasars.

## ACKNOWLEDGMENTS

The author would like to thank the anonymous referee for very helpful comments and suggestions.
This research has made use of the NASA/IPAC Extragalactic Database (NED), which is operated by the Jet Propulsion Laboratory, California Institute of Technology, under contract with the National Aeronautics and Space Administration.

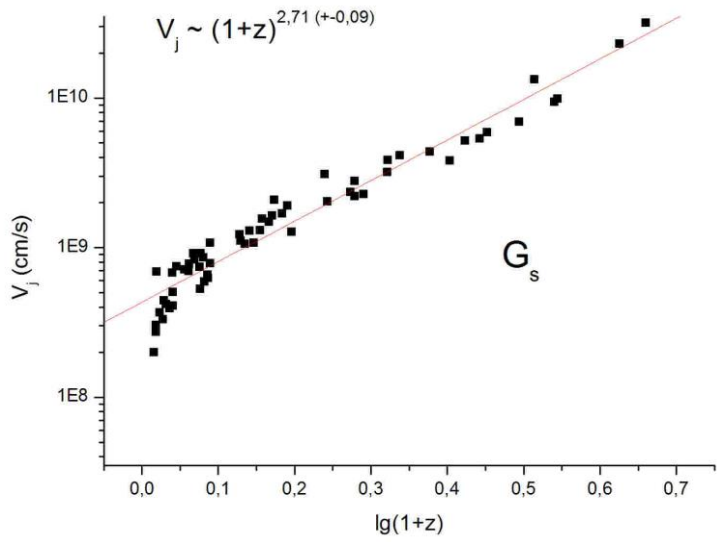

Figure 2. Jet propagation velocities of galaxies with linear steep spectra versus redshift.

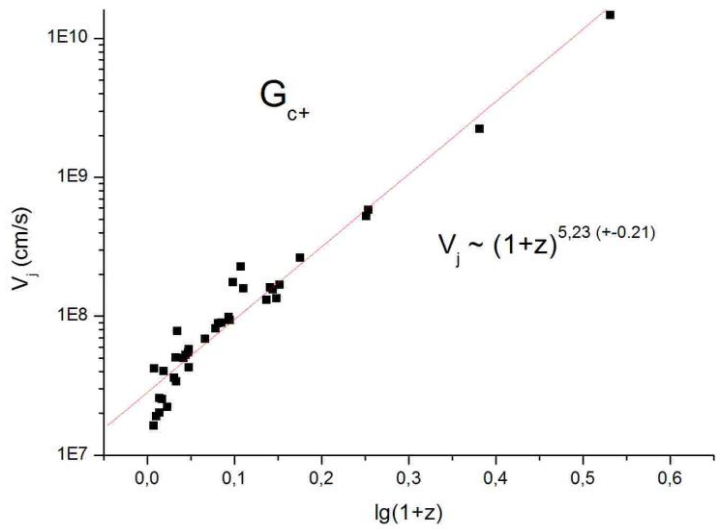

Figure 3. Jet propagation velocities of galaxies with break steep spectra versus redshift.

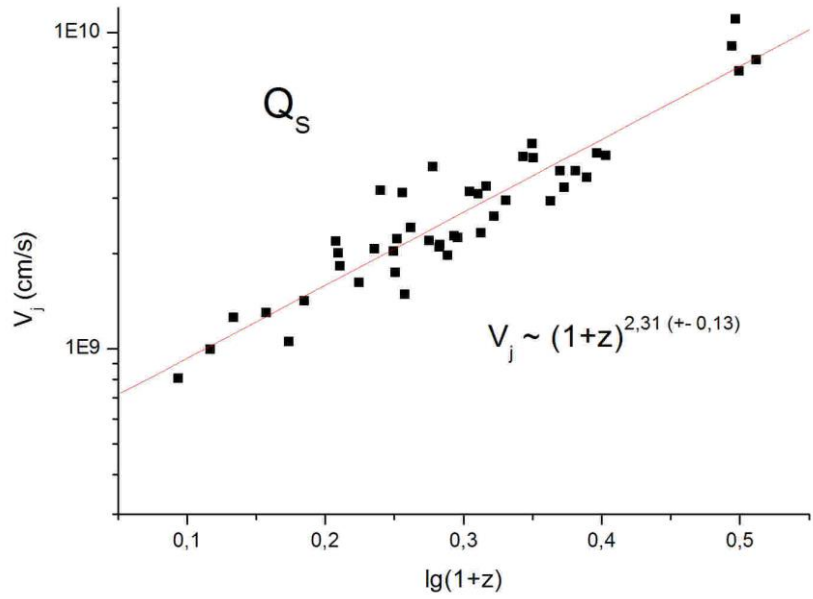

Figure 4. Jet propagation velocities of quasars with linear steep spectra versus redshift.

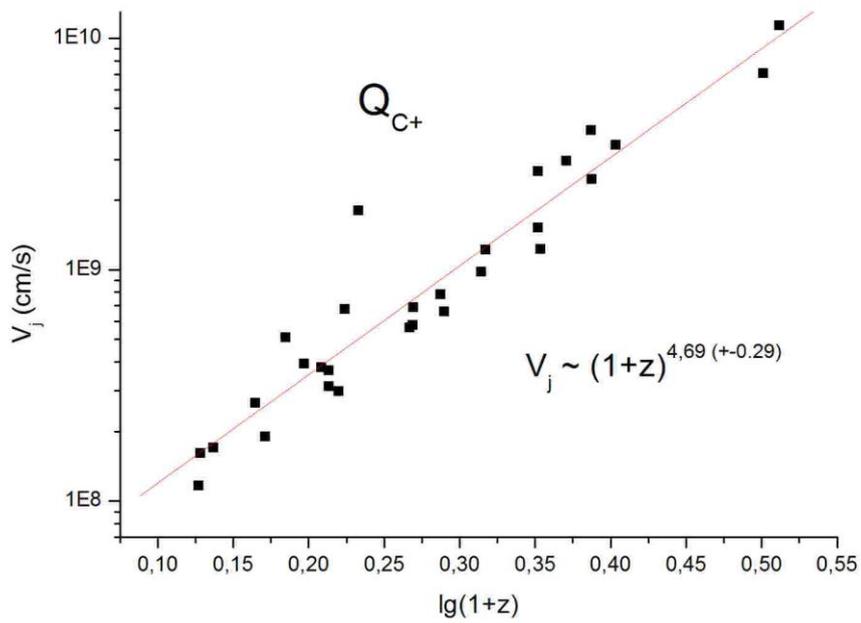

Figure 5. Jet propagation velocities of quasars with break steep spectra versus redshift.

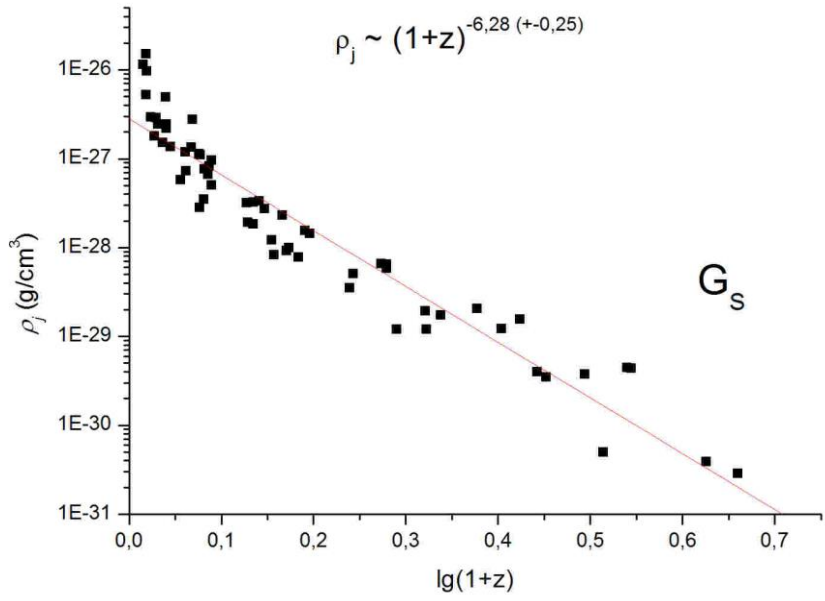

Figure 6. Jet environment densities of galaxies with linear steep spectra versus redshift.

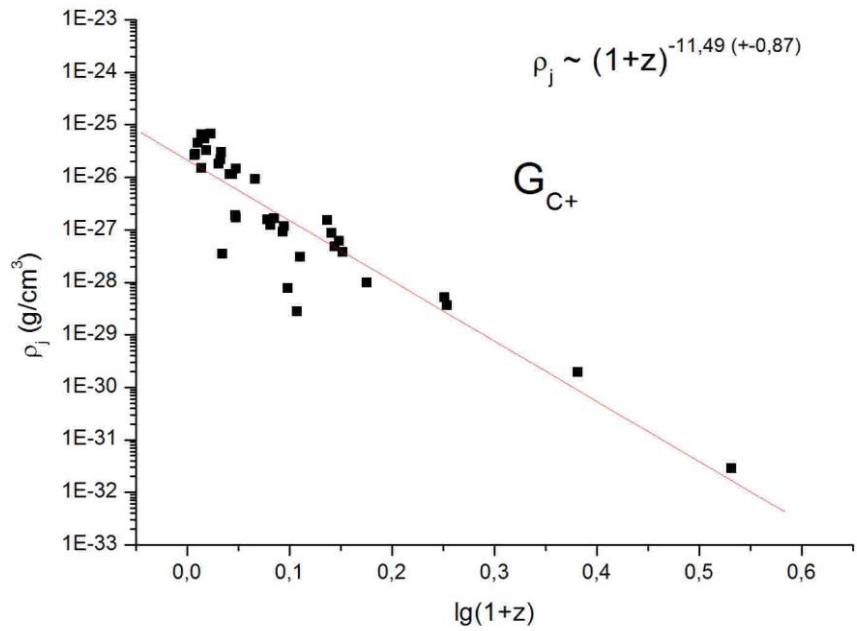

Figure 7. Jet environment densities of galaxies with break steep spectra versus redshift.

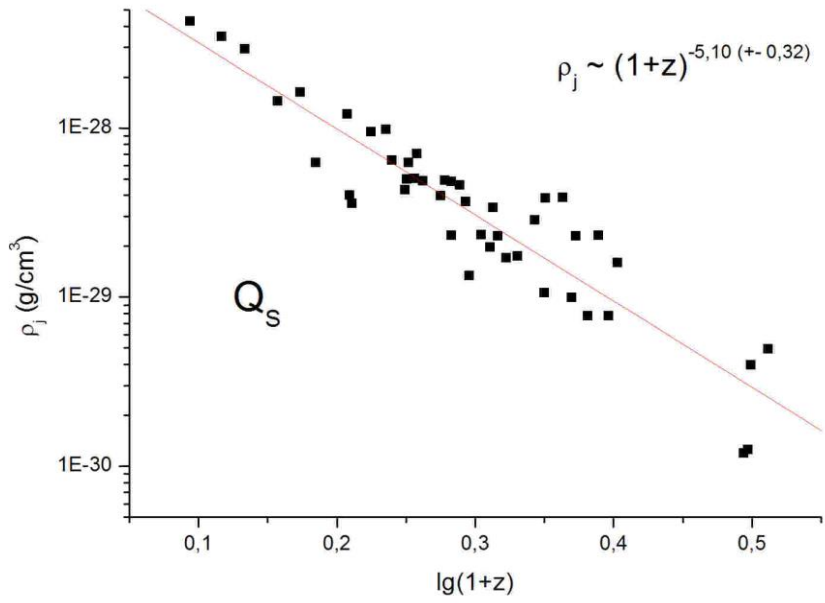

Figure 8. Jet environment densities of quasars with linear steep spectra versus redshift.

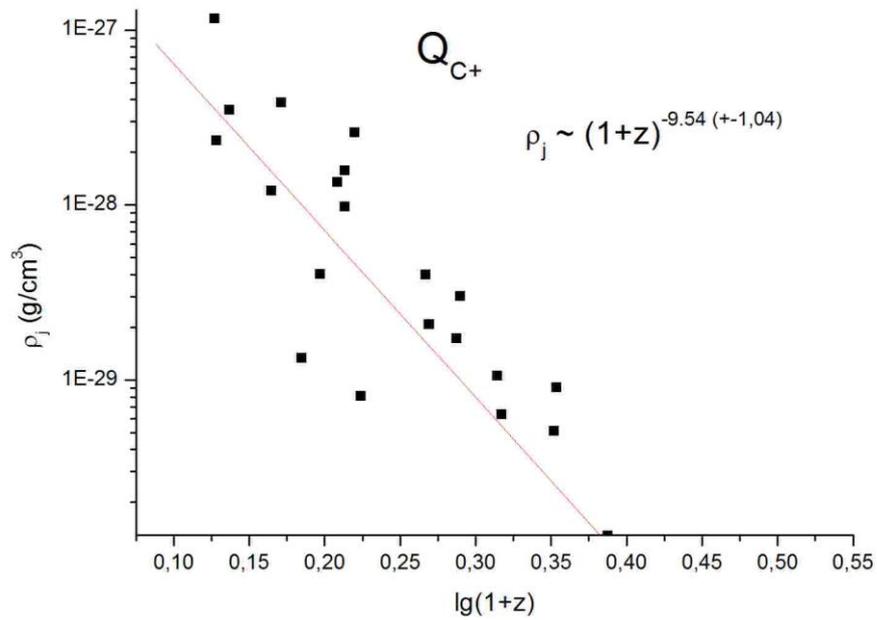

Figure 9. Jet environment densities of quasars with break steep spectra versus redshift.